# EMPLOYEE TRUST BASED INDUSTRIAL DEVICE DEPLOYMENT AND INITIAL KEY ESTABLISHMENT

Apala Ray[1, 2] and Johan Akerberg[2, 3] and Mats Bjorkman[3] and Mikael Gidlund[4]

[1] ABB Corporate Research, Bangalore, India
[2] Malardalen University, Vasteras, Sweden
[3] ABB Corporate Research, Vasteras, Sweden
[4] Mid Sweden University, Sundsvall, Sweden

## ABSTRACT

*An efficient key management system is required to support cryptography. Most key management systems use either pre-installed shared keys or install initial security parameters using out-of-band channels. These methods create an additional burden for engineers who manage the devices in industrial plants. Hence, device deployment in industrial plants becomes a challenging task in order to achieve security. In this work, we present a device deployment framework that can support key management using the existing trust towards employees in a plant. This approach reduces the access to initial security parameters by employees; rather it helps to bind the trust of the employee with device commissioning. Thus, this approach presents a unique solution to the device deployment problem. Further, through a proof-of-concept implementation and security analysis using the AVISPA tool, we present that our framework is feasible to implement and satisfies our security objectives.*

## 1. INTRODUCTION

Industrial control systems, which include Supervisory Control and Data Acquisition (SCADA) systems, Distributed Control Systems (DCS), and Programmable Logic Controllers (PLC), are used to monitor and control industrial processes. These control systems acquire data from an industry process for monitoring and issue control commands whenever required. Industrial control systems are typically used in process industries like pulp and paper, water and wastewater, food and beverages, mining etc. A typical paper mill can have thirty to fifty thousand sensors and actuators. The goal of industrial automation is to automate the operations involved in the technical process with minimal or reduced human intervention. In the initial phase of industrial automation, industrial plants were built as stand-alone systems, where specialized hardware and software were used by proprietary control protocols. Many of these components were not connected with the outside world, so security had less attention. Since the last decade, industrial communication security has gained a lot of research interest. The reason is that companies start to introduce Internet in a larger extent than before. This has posed the possibility of cyber threats in industrial segments. Communication security with security objectives, types of attack, cryptographic methods, security in communication protocols and security best practices is discussed in [1]. The industrial communication security aims to protect the devices (sensors/actuators/controllers) from any kind of security attacks. The security attacks from the IT domain are also affecting the industrial automation domain. Recent known attacks like Stuxnet have revealed another set of challenges where malware can spread itself, for example through USB drives, and when finding the target system it can infect the PLC's with a Trojan [2]. In the security domain, cryptography is a well-known technique to protect communication between devices from attackers. Generally, different cryptographic algorithms are used for communication security and the security of cryptographic algorithms relies on underlying secret parameters. To create a secure system, the initial setup for the cryptography details is very important. Therefore, an effective key management in industrial plants is an important requirement for having a secure system.

## 1.1. Motivations

Industrial plants have specific requirements on availability and at the same time on easier workflow for the commissioning and maintenance engineers. The explicit assumption to have a secured system is that the devices in the network are trusted. This trust may be established by the explicit mechanism of out-of-band initial trust bootstrapping, such as manual entry of security key parameters in the device. The issues involved in the assumptions or pre-requisite of "key distribution" are discussed in detail in [3]. For instance, considering the large number of devices inside a plant, such out-of-band initial trust bootstrapping methods create an additional burden for engineers. It is also a non-trivial task for a commissioning and maintenance engineer to find the physical devices that are spread over large areas and to configure with the right parameters for each of the devices without transmitting secret keys.

Industrial plants also involve many employees for successful operation of the plant. Each employee has a specific role in managing the plants for 24x7 operations. There are the following roles relevant to security management in industrial plants, (a) manufacturers of the devices, (b) system integrators who customize the devices, integrate them into the plant and perform commissioning, (c) operators who monitor the system during their normal operation and respond to alarms, and (d) service personnel who are responsible for maintaining and repairing the devices [4]. In addition to this, these roles might be manned from different organizations. For example, the system integrators of the plant may be the manufacturer, the asset owner, or an external company. These roles are involved in operation of the plants including the device functionality and their management. Successful function of plants is possible when the devices are properly commissioned, operated and maintained. Therefore, the security management of devices inside the plant is indirectly coupled with the different employees and their roles. The device management can be restricted based on a role-based access control policy [5]. However, there might be several employees who share the same role. For instance, in a medium size plant, there might be fifty employees who are assigned to commission the plant. Therefore, the role-based access control cannot guarantee accountability for an individual employee in case of device configuration.

For a successful security deployment in the plant, it is necessary to create accountability and establish a relationship between the employee and the device. At present, the industrial automation life-cycle does not have a workflow which can link and manage both the device security and the employee access. Therefore, there is a need to harmonize the link between device security and employee access. An idea of distributing the initial trust to the devices in a comparatively simple workflow for the commissioning and maintenance engineers is proposed in [6]. In this paper, we enhance the idea of integrating the responsibility of an employee management system with the security management component for the device management in the plant.

## 1.2. Contributions

In this paper, we present an industrial device deployment framework based on the initial bootstrapping of trust from employees.

- We propose a framework to logically segregate the feature of security management of devices from the role of employees in a plant. This independent device deployment framework considers the dynamic environment of employee's roles in industrial plants.
- We also propose a mechanism for the device to verify whether it is joining the intended network.
- We also propose key generation and key deployment mechanisms for heterogeneous types of plants with devices of varying degree of computation capabilities.
- Our framework is also adaptive and can be used where the devices do not have direct connectivity with the central security management or employee management system.
- Through a proof-of-concept implementation and security analysis, we show that the proposed framework is feasible to implement and satisfies the security objectives.
- We also simulate the proposed schemes and methods using the AVISPA (Automated Validation of Internet Security Protocols and Applications) tool to validate the protocols used in the framework.

## 1.3. Paper Structure

In this paper, section 2 discusses the related work. Section 3 presents an overview of the proposed framework of industrial device deployment along with the trust and the threat model. In section 4, the framework is discussed in detail. Section 5 presents the details of the proof-of-concept implementation. The assessment of our proposed framework is presented in section 6. Finally, conclusions are presented in section 7.

## 2. RELATED WORK

There is extensive and ongoing work on topics addressing key management issues. A. Kumar et al. presented a detailed survey on the key management protocols for wired and wireless networks [7]. S. Camtepe covers deterministic, probabilistic and hybrid pre-distribution schemes for distributed networks and propose to establish pair-wise, group-wise and network-wise keys in hierarchical networks [8]. This work analyzes many of the security and efficiency related characteristics. Generally there is no single solution which can solve all key distribution related problems. Additionally, in each of the key distribution approaches, there is either an explicit assumption or an explicit mechanism to establish the initial parameters among the communication parties. The explicit assumption is that the devices in the network are trusted or there is an explicit mechanism of out-of-band parameters sharing. K. Fischer et al. compare different approaches to initially bootstrap security credentials [9]. In this work, the authors concluded that the best method to bootstrap initial credentials can be done through manufacturer provided certificates. The automation device is manufactured by the device vendor and equipped with a secure device identifier based on 802.1 AR [10]. However, this imposes a tight constraint on manufacturers to provide a device with secure device identity. This also might increase the manufacturing effort and costs as the credential generation will be included during production process. F. Stajano et al. [11] discussed the issues of bootstrapping security devices and proposed an solution to configure the trust relation of a device with a help of users. However, their solution requires physical contact of the new device with a master device and the new device stay loyal to master device. A. Perrig et al. present a special way of key distribution based on a master-key pre-loading approach [12]. However, it needs to setup a shared secret key between each node and the base station, as a pre-requisite for key distribution. L. Eschenauer et al. proposed a key management scheme using probabilistic key sharing [13], which was improved by C. Haowen and W. Du et al. [14, 15]. F. Gandino et al. proposed a random seed distribution with transitory master key [16]. However, these type of schemes also need offline loading of keys prior to distribution. A concept of polynomial key pre-distribution based on deployment knowledge is presented by D. Liu et al. [17, 18]. Using deployment knowledge, a key pre-distribution concept based on a key pool has been shown by Z. Yu et al. [19]. However, these mechanisms have pre-requisite that each group of nodes should share the same secret matrix. Using this matrix, pairwise keys can be generated between nodes. M. Shehab and V. Bulusu et al. presented a hierarchical key distribution for sensor networks [20, 21]. K. Xue et al. presented security improvement of a hierarchical key distribution mechanism for large-scale Wireless Sensor Network [22] which was proposed by Y. Cheng et al. [23]. These schemes require pre-loading of a `polynomial share' within the nodes before deployment. A secure and efficient network bootstrapping protocol for 6LoWPAN has been proposed by H. Cha et al. [24], where challenge response mechanism can be used for secure joining. However, this does not cover the initial credential distribution process for authentication. Flaws of single-sign-on schemes are discussed by G. Wang et al. [25]. There has been some research work using the advantage of multi-path signal propagation as a source of randomness to generate secrets [26-28]. M. Wilhelm et al. showed a key deployment protocol using key generation from physical layer information [29]. This provides an elegant and user-friendly mechanism to the key deployment problem; however the capability of generating ephemeral shared secrets from industrial channel measurements needs to be verified. A tamper-evident pairing protocol that provides simple, secure Wi-Fi pairing and protects against Man-In-The-Middle (MITM) attacks without an out-of-band channel has been shown by S. Gollakota et al. [30]. This is an interesting solution for Wi-Fi devices with push button configuration. It does not require out-of-band key pre-distribution, however it requires pressing of push button on the Wi-Fi devices for initiating the mechanism. Smart card based authentication is also discussed by J.-L. Tsai et al. [31]. An assessment of the current security situation of industrial distributed computing systems has been discussed by M. Cheminod [32]. The authors believe that because of the complexity and size of many industrial plants, quick and effective security management decisions and (re)actions will become harder to take in the near future, so that the scientific community is expected to propose and develop new advanced techniques. The LTE security is explained by D. Forsberg et al. in detail [33]. The SIM card or certificate based solutions in mobile telecommunication industry require a lot of engineering either in manufacturer premises or in the industrial plant itself. A SIM card based solution requires individual mapping between the SIM card and the devices, which adds extra time consuming steps in the industrial workflow.

From the related work and to the best of our knowledge, there is no automated workflow of initial credential distribution solution for industrial devices. There is either an assumption or a pre-requisite of initial key availability in the industrial devices prior to the secure key distribution. In industrial plants, employees manage devices, and the employees can be identified through their registered identity with the system. Therefore, in this work we propose a workflow to use the already established trust of the employees for enabling the initial bootstrap of trust in the devices. The flexibility of this approach enables commissioning engineers to download the required configuration data in the device. This approach is a unique solution to the initial trust distribution problem reusing the existing features and facilities in industrial plants.

# 3. SYSTEM ARCHITECTURES, THREAT MODEL AND SYSTEM OBJECTIVE

In this section, we present our proposed concept and the design goals of a device deployment framework for industrial plants. We also describe the components we need to use in this framework along with the assumptions. The initial trust of the employee is transferred to the device during the commissioning phase of the plant life-cycle and we assume that this step can be performed either by the manufacturers, the asset owners, or external companies. In our framework, the employee management system keeps track of physical accesses for all the employees where they are authorized to enter in the different areas and rooms in the plant, as well as handling the devices. Furthermore, the plant also has a security management component to handle the security of the devices.

## 3.1. System Architecture:

### 3.1.1. System components and Trust Model

The components which are used in the device deployment framework are presented below. In Table 1, we summarize the trust assumptions for the system components.

- *Security management component*: This component handles the security parameters required for the device communication, and monitors the security state of the devices in a running plant. This component has to be the most secure component as it will be the weakest link in the security chain. If this component is compromised, then the security chain will be broken. If there is any other security management system within the plant, this component will coordinate with that system.

- *Employee management system:* This component is responsible for issuing ID cards to employees. At the plant there is physical security and a first level of access control is used to securely store the employee access data. The employee might be from an organization such as the manufacturer, site owner, or a third party. The details of the employees who are going to handle the devices are stored in this component.

- *Commissioning engineer/maintenance engineer:* This engineer is authorized to configure or commission devices prior to the operational phases or during the maintenance phase. The employee has an identity card which is registered with the *Employee management system*. A unique password for the identity card is required and this password is the same password which is used to get physical access to the building.

- *ID card of a commissioning engineer:* The information related to the *Commissioning engineer/maintenance engineer* provided by the *Employee management system* is stored inside this component. This component is used for transferring the trust of the engineer to the devices.

- *Commissioning device:* This component is primarily used as a medium for transferring the trust of engineer to the device.

- *Slave device:* This component is the device which needs access for the network. During the commissioning phase, the trust from the commissioning engineer is transferred to this component.
- *Master device:* This component resides at the upper communication level from the *Slave device*.

Table 1.  Notations used in Deployment Framework.

| Components | Trust Assumption |
|---|---|
| Security management component | This component cannot be compromised |
| Employee management system | This component cannot be compromised |
| Commissioning engineer/ maintenance engineer | This person is trusted from the organization and keeps the own password confidential. Reporting the loss of the identity card is expected from this person |
| ID card of a commissioning engineer | The content of this card can only be accessed through the employee password |
| Commissioning device | This is a trusted component in the plant. When it reads the content of the card through employee password, it stores it in temporary memory. When the information is properly transferred to the device, it erases the content immediately |
| Slave device | The trust assumption is similar to current industrial devices where physical access control is present for field devices. Firmware analysis or side channel attacks are not possible when the device is commissioned inside the plant |
| Master device | The trust assumption is similar to current industrial devices where physical access control is present for field devices. Firmware analysis or side channel attacks are not possible when the device is commissioned inside the plant |

### 3.1.2. Threat Model

The adversary is an ordinary device or a resourceful device which can create malicious activities in a network. This threat model defines adversaries and their possible attacks to the proposed framework. We focus on proposing a framework which can mitigate the threats which can arise from this threat model.

- Adversaries can listen to message exchanges between slave device, master device and security manager
- Adversaries can inject messages in the network
- Adversaries can capture or replay messages later
- Adversaries can steal the ID card of an employee

### 3.1.3. Framework Overview

The proposed device deployment framework consists of basically three phases as shown in Figure 1, which presents a simplified conceptual overview of our proposed industrial device deployment framework.

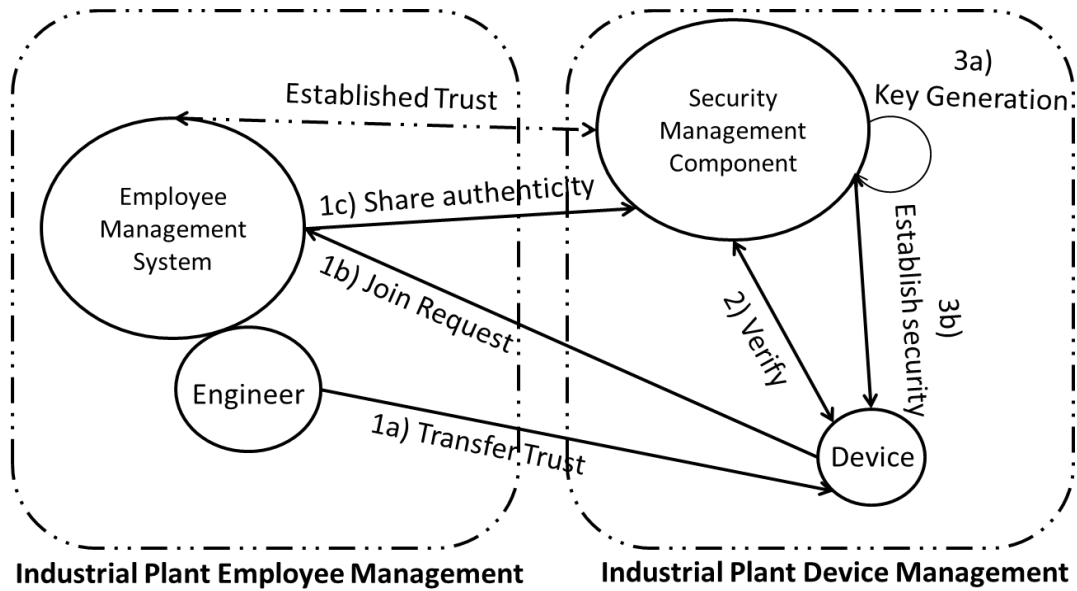

Figure 1. Device Deployment Framework

**Initial Trust based authentication:** In the first phase, the initial trust is established where the commissioning engineer/maintenance engineer configures the device and the trust of the engineer is transferred to the device. The device capabilities can also be stored into the device during commissioning. The device is authenticated based on the trust of the engineer which was transferred to the device during commissioning.

**Authenticity Verification:** In the second phase, the device is verified whether it can present the proof of possessing the correct trust information. The device also verifies whether it is joining the intended network.

**Key Establishment:** In the third phase, the key generation occurs for the device. Based on the device capability, the security management component decides which type of key should be generated for the device. In a plant, there are different types of devices with different computational resources. Our framework is designed for such heterogeneous types of systems. Therefore, based on the device capabilities, the asymmetric keys or a symmetric key is generated by the security management component. These keys can either be used for secure single-hop communication, or to support end-to-end encryption in multi-hop topologies. If the device is capable of generating its own key, it can share its key with the security management component once the verification phase is done.

The proposed framework is developed to support hierarchical trust establishment. In this framework, some of the devices might have direct connectivity with the employee management system and can be directly verified by the employee management system. We define these devices as Level 1 trusted devices. Once the trust relation is established between the employee management system, the security management component and the Level 1 devices, these Level 1 devices can be used to anchor the trust establishment procedure for next level devices. The next level devices will have one-hop connectivity with the employee management system. In our proposed framework we categorize the initial trust establishment in two scenarios. In the first scenario as captured in Figure 2, the device can be directly verified by the employee management system. We define the first scenario as *direct topology*.

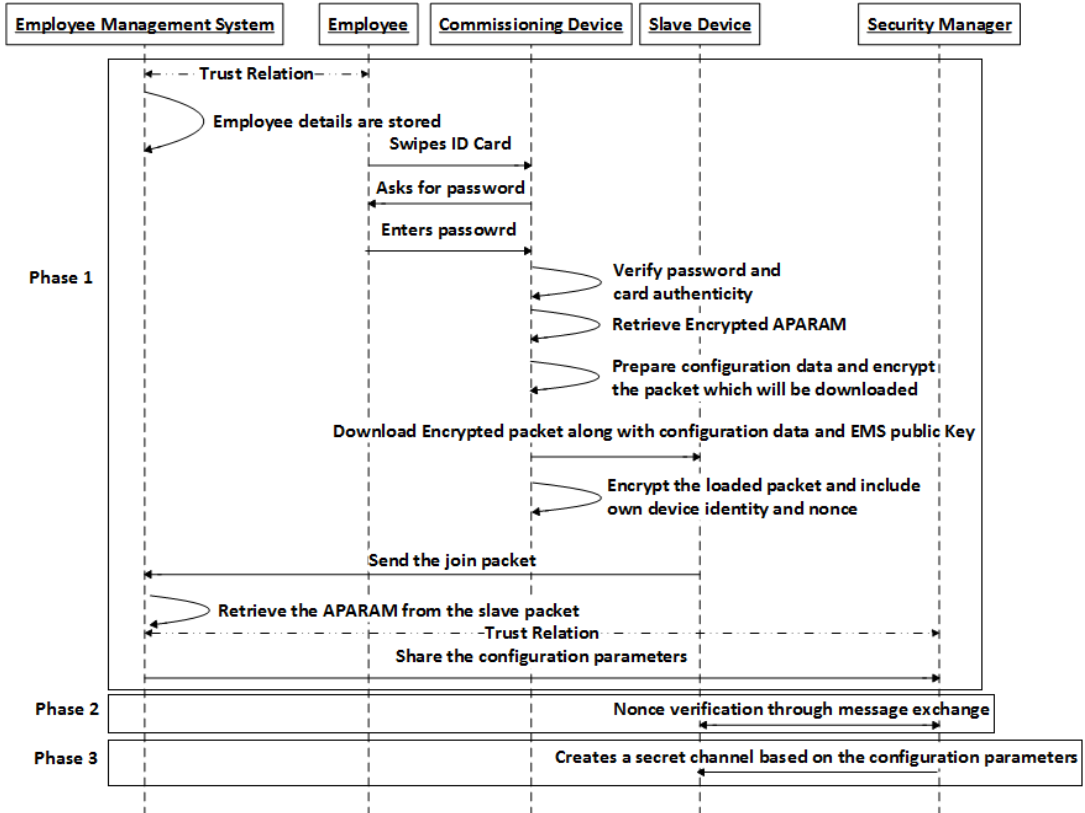

Figure 2.  Device Deployment – Direct Topology

In the second scenario as captured in Figure 3, the device can be verified by employee management system through an intermediate device, such as a master device. We define the second scenario as *hierarchical topology*.

## 3.2. System Objectives

The security objective of industrial communication is to ensure that all the entities in the industrial plants are communicating through a secure channel. This implies that the plant is required to have an infrastructure where devices are deployed and the secure communication channel is established. This leads to an efficient security management scheme for industrial environments. Our proposed framework is designed to meet the following identified objectives. The framework is also supposed to maintain the basic properties of crypto for confidentiality, integrity and device authentication.

*Initial secret key never leaves the node:* The security parameters which will be shared between two devices should stay within devices, such that only intended devices can read the parameters. *System resilience:* Compromise of one device should have minimal impact on the rest of the system.

*Accountability for device configuration:* The person who has configured the device should be traced.

*Ease of configuration:* Replacing or adding a device should be easier for any employee without having in-depth security understanding.

*Time to configure:* The system should allow fast access to devices for replacement or extension by the authorized users.

*Ease of system deployment:* This property demands that the workflow can be deployed without much effort to set up or maintain the security life-cycle.

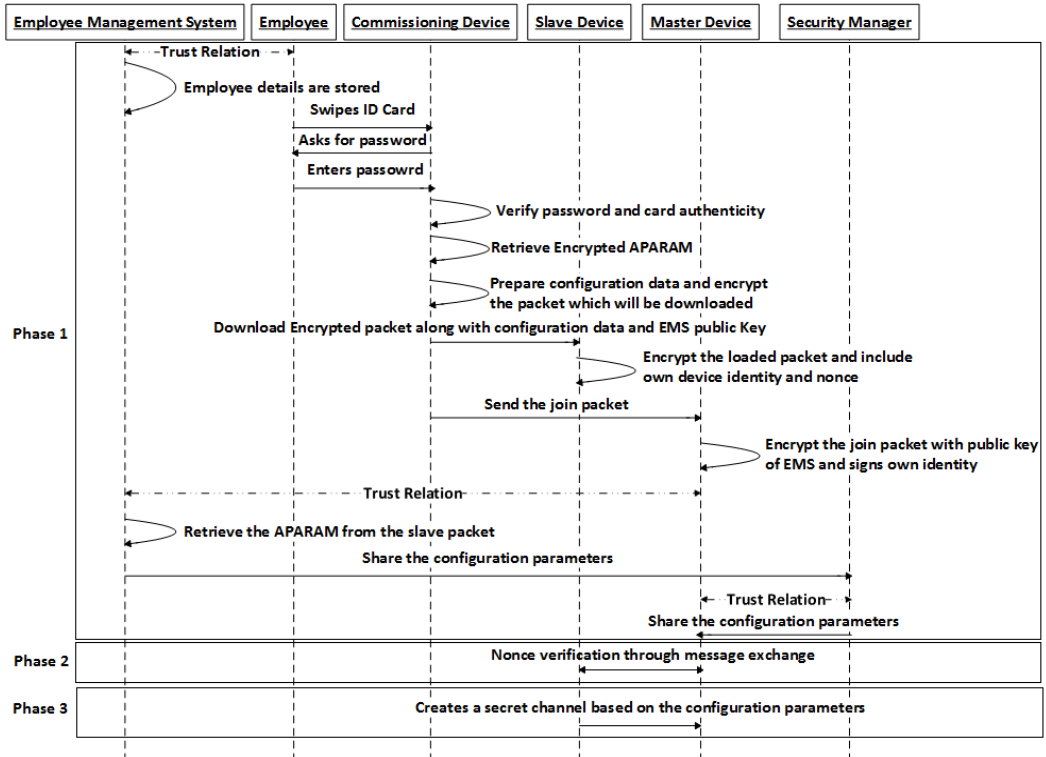

Figure 3. Device Deployment – Hierarchical Topology

# 4. INDUSTRIAL DEVICE DEPLOYMENT – FRAMEWORK

This section introduces the industrial device deployment framework with the security protocols in more detail. Our proposed algorithms are used in one time activity for bootstrapping. In Table 2, we summarize the notation used in the framework description to make easier for readers to refer to.

Table 2. Notations used in Deployment Framework.

| A → B :< M > | A sends message M to B |
|---|---|
| *ID* | ID card of the Commissioning Engineer |
| *HH* | Commissioning Device Handheld |
| *S* | Slave Device |
| *M* | Master Device |
| *EMS* | Employee Management System |
| *SM* | Security Management Component |
| *EMP* | Employee Commissioning Engineer |
| $A_{ID}$ | Unique identity of any device A |
| $sign(A_{ID})$ | Signature of any device ID $A_{ID}$ |
| *E(K,T)* | Encryption function for text T with key K |
| *D(K,T)* | Decryption function for text T with key K |
| *inc(N)* | Increment function for N |

| $Cert_{EMS}$ | Certificate of $EMS$ |
|---|---|
| $K_{pr}(A)$ | Private key of any device A |
| $K_{pub}(A)$ | Public key of any device A |
| $K_{A-B}$ | Symmetric key between device A and B |
| $NONCE_S$ | Random numbers generated by slave to prevent message replay and support authentication verification |
| $RND_A$ | Random numbers generated by any device A to support authentication verification |
| $APARAM$ | Authentication parameter for employee |
| $ENC_{APARAM}$ | Authentication parameter encrypted with $EMS$ public key |
| $CD$ | Configuration Data Packet commissioned for slave device |
| $P_{authComm}$ | Encrypted configuration Packet along with authentication parameter downloaded to slave |
| $P_{join}$ | Joining Packet sent from slave to master or $EMS$ |
| $P_{joinFwd}$ | Forwarded Joining Packet for slave sent from master to $EMS$ |
| $P_{authDev}$ | Authorized Packet for device sent from $EMS$ to $SM$ |
| $P_{DH}$ | Packet shared between slave to master or $EMS$ during DH |
| $a,b$ | Large random numbers used in the DH key exchange |
| $A,B$ | Public keys used in the DH key exchange |
| $K_S$ | Key used between slave and master or slave and $SM$ based on DH |

## 4.1. Initial Trust based authentication phase

At the beginning of the initial trust setup phase, the commissioning engineer or maintenance engineer swipes the *ID card* in the commissioning device *HH* and enters the password. The encrypted authentication parameters $ENC_{APARAM}$ is stored in the *ID card*. The *HH* verifies the password and the $ENC_{APARAM}$ with the $EMS$ certificate $Cert_{EMS}$.

Once this verification is done, the *HH* creates a packet with the configuration data *CD* and the $ENC_{APARAM}$. The *CD* may contain the identity of the commissioning engineer $EMP_{ID}$ and optionally the identity of the commissioning device $HH_{ID}$ along with the device configuration details. As a next step, the *HH* encrypts the *CD* and $ENC_{APARAM}$ with the public key of the employee management system $K_{pub}(EMS)$. This encrypted packet is denoted as $P_{authComm}$. Then $P_{authComm}$ along with the *CD* and the $K_{pub}(EMS)$ are downloaded in the device (*S*). The $P_{authComm}$ can also be stored in tamper proof memory of the device, so that if the device is captured by the adversary, the information cannot be retrieved from the device.

In the initial trust based authentication as shown in Algorithm 1, the slave device *S* generates one random nonce $NONCE_S$. It also appends its own device identity $S_{ID}$ and then it encrypts the downloaded $P_{authComm}$, $S_{ID}$, and $NONCE_S$ with the $K_{pub}(EMS)$. This encrypted packet is denoted as $P_{join}$. The $P_{join}$ is sent to the higher level devices for further security management.

In *direct topology*, the slave device *S* has direct connectivity with the employee management system. The employee management system can retrieve the content of the packet $P_{join}$ using the private key of the employee management system $K_{pr}(EMS)$. It retrieves $P_{authComm}$, nonce and slave device identity. Then again using the $K_{pr}(EMS)$ it retrieves the encrypted authentication parameter and then after another decryption, it retrieves $APARAM$. This authentication parameter $APARAM$ can only be downloaded by an authorized engineer having an authenticated ID Card. Therefore, through the secret $APARAM$ within the packet $P_{join}$, the employee management system can verify that the device is commissioned by an authorized person. The employee management system has a trusted connection with the security management component *SM*. *EMS* signs its own identity $EMS_{ID}$ with $K_{pr}(EMS)$ and create packet $sign(EMS_{ID})$. It also creates a packet $P_{authDev}$ by encrypting the $CD$, the $NONCE_S$ and the $sign(EMS_{ID})$ with the public key of the security management component $K_{pub}(SM)$. Then the employee management system sends the packet $P_{authDev}$ to the security management component.

---

## Algorithm 1 Initial Trust Based Authentication

1: **procedure** INITIALTRUSTAUTH

$\quad ID \rightarrow HH : \quad < ENC_{APARAM} >$

$\quad HH : \quad P_{authComm} = E(K_{pub}(EMS), (CD, ENC_{APARAM}))$

$\quad HH \rightarrow S : \quad < K_{pub}(EMS), P_{authComm}, CD >$

$\quad S : \quad NONCE_S = random()$

$\quad S : \quad P_{join} = E(K_{pub}(EMS), (P_{authComm}, S_{ID}, NONCE_S))$

2:    **if** Direct topology **then**

$\quad S \rightarrow EMS : \qquad\qquad < P_{join} >$

3:    **else if** Hierarchical topology **then**

$\quad S \rightarrow M : \quad < P_{join} >$

$\quad M : \quad sign(M_{ID}) = E(K_{pr}(M), M_{ID})$

$\quad M : \quad P_{joinFwd} = E(K_{pub}(EMS), (P_{join}, sign(M_{ID})))$

$\quad M \rightarrow EMS : \quad < P_{joinFwd} >$

$\quad EMS : \quad P_{join}, sign(M_{ID}) = D(K_{pr}(EMS), P_{joinFwd})$

4:    **end if**

$\quad EMS : \quad P_{authComm}, S_{ID}, NONCE_S = D(K_{pr}(EMS), P_{join})$

$\quad EMS : \quad CD, ENC_{APARAM} = D(K_{pr}(EMS), P_{authComm})$

$\quad EMS : \quad APARAM = D(K_{pr}(EMS), ENC_{APARAM})$

$\quad EMS : \quad sign(EMS_{ID}) = E(K_{pr}(EMS), EMS_{ID})$

$\quad EMS : \quad P_{authDev}$
$\qquad\qquad = E(K_{pub}(SM), (CD, NONCE_S, sign(EMS_{ID})))$

$\quad EMS \rightarrow SM : \quad < P_{authDev} >$

5: **end procedure**

---

In *hierarchical topology*, the slave device does not have direct connectivity with the employee management system. Therefore, in that case, the slave device $S$ sends the packet to the master device $M$. Master device signs its identity $M_{ID}$ with the private key of the master device $K_{pr}(M)$ and creates the packet $sign(M_{ID})$. Using authentication of direct topology, master device has already established the trust relation with the security management component and the employee management system, it encrypts the packet $P_{join}$ and $sign(M_{ID})$ with the public key of the employee management system $K_{pub}(EMS)$ and sends the encrypted packet $P_{joinFwd}$ to the employee management system. The employee management system decrypts the packet $P_{joinFwd}$ with its private key $K_{pr}(EMS)$ and retrieves $P_{authDev}$ and $sign(M_{ID})$. Then it can verify the identity of the master device $M_{ID}$ through the public key of the master device $K_{pub}(M)$ and can retrieve the content of the packet $P_{authComm}$ using the private key of the employee management system $K_{pr}(EMS)$. Once the employee management system can verify that the master device has forwarded the data from a slave device which is commissioned by a trusted person, it shares the information of the slave device with the security management component in a similar way as in the case of direct topology.

## 4.2. Authenticity verification phase

The goal of this phase is to ensure that the device which presents the trust information from the employee can also present the proof of possessing the correct trust information before establishing the key between the device and the security management component. At the same time, the device should also ensure that it is joining the correct network which it is supposed to join. As shown in Algorithm 2, the security management component can retrieve the content of the packet $P_{authDev}$ which is forwarded by the employee management system, using the private key of the security management component $K_{pr}(SM)$.

---

**Algorithm 2** Authenticity Verification Phase

1: **procedure** AUTHVERIFICATION

   $SM:$      $CD, NONCE_S, sign(EMS_{ID}) = D(K_{pr}(SM), P_{authDev})$

2:    **if** Direct topology **then**

   $SM:$      $NONCE_{SM} = random()$

   $SM \rightarrow S:$      $< E(NONCE_S, (NONCE_{SM}, inc(NONCE_S))) >$

   $S:$      $RND_S = random()$

   $S \rightarrow SM:$      $< E(NONCE_{SM}, (RND_S, inc(NONCE_S))) >$

3:    **else if** Hierarchical topology **then**

   $SM:$      $sign(SM_{ID}) = E(K_{pr}(SM), (SM_{ID})) >$

   $SM \rightarrow M:$      $< E(K_{pub}(M), (NONCE_S, sign(SM_{ID}))) >$

   $M:$      $RND_M = random()$

   $M \rightarrow S:$      $< E(NONCE_S, (NONCE_M, inc(NONCE_S))) >$

   $S:$      $RND_S = random()$

   $S \rightarrow M:$      $< E(NONCE_M, (RND_S, inc(NONCE_S))) >$

4:    **end if**
5: **end procedure**

---

During authentication in *direct topology*, the security management component generates a random number $RND_{SM}$ and increments the nonce $NONCE_S$ by 1. Then it sends the packet to the slave by encrypting it with the $NONCE_S$. The slave device can decrypt the content as it has the generated nonce $NONCE_S$ and read the $RND_{SM}$ and incremented $NONCE_S$. Thus, the slave knows that the packet has come from an authorized component that has retrieved the correct configuration data from the slave. The slave device again generates a random number $RND_S$ and increments the incremented $NONCE_S$ by 1, then it encrypts the $RND_S$ and $inc(NONCE_{S)}$ with $RND_{SM}$. Once the security management component gets this new packet from the slave, it can verify that the slave device possesses the correct configuration data as it was configured by an authorized engineer.

In *hierarchical topology*, the security management component signs its own identity and create $sign(SM_{ID})$. Then it encrypts the $NONCE_S$ and $sign(SM_{ID})$. With the public key of the master device $K_{pub}(M)$. Here, the assumption is that the master device can support public key cryptography. If the master device does not support public key cryptography, then the packet can be encrypted with the common shared key between the security management component and the master device. The rest of the verification phase to verify whether the slave device possesses the correct configuration data is similar to direct topology.

## 4.3. Key establishment phase

The goal of this phase is to establish an authenticated secret which will be used to protect the communication in the network. In our framework, we have focused to bootstrap the device trust so that key management can be done from a centralized component. Once the devices are verified inside the plant as properly commissioned by an engineer, then the security manager component can enforce the key establishment for the network as different state-of-the-art key establishment.

### 4.3.1. Symmetric Key based security management

As shown in Algorithm 3, during authentication in *direct topology*, both the security management component and the slave device will use the same key if symmetric key based security management is used. In hierarchical topology, both the master device and the slave device will use a common key.

**Algorithm 3** Symmetric key establishment

1: **procedure** SYMKEYGEN
2:    **if** Direct topology **then**

     $SM:$                 $K_{SM-S} = KeyGen()$

     $SM \rightarrow S:$       $< E(RND_S, (K_{SM-S}, inc(NONCE_S))) >$

3:    **else if** Hierarchical topology **then**

     $M \rightarrow S:$        $< E(RND_S, (K_{M-S}, inc(NONCE_S))) >$

4:    **end if**
5: **end procedure**

In *direct topology*, the slave device has direct connectivity with the employee management system and once the device is verified, the security management component generates the key $K_{SM-S}$ which will be used for first time communication between the security management component and the slave device $S$ and is later replaced by the security manager component which enforces standard key establishment for the network as state-of-the-practice. Then it encrypts the $K_{SM-S}$ with $RND_S$ and sends it to the slave device.

In *hierarchical topology* once the device is verified by master device, the master device uses the key $K_{M-S}$ which can be received from the security management component or it can be generated by the master device if the master device has the key generation capability. It then encrypt $K_{M-S}$ with $RND_S$ and send encrypted $K_{M-S}$ to the slave device.

**4.3.2. Asymmetric Key based security management**

If devices have the necessary computation power for public key cryptography operations once in a while, then this workflow will be suitable for those types of devices. This concept is similar to Password-based Encrypted Key Exchange [34].

As shown in Algorithm 4, the security management component generates secret key a and computes $A = g^a \bmod p$. The modulus $p$ and base exponent $g$ are the parameters denoted as $PUB_{DH}$. Then the security management component increments the nonce by 1 and creates the packet with $A$, $PUB_{DH}$ and nonce. It then forwards the packet to the slave device encrypting with $RND_S$. The encrypted packet is denoted as $P_{DH}$. The slave device decrypts the packet $P_{DH}$ with $RND_S$ and retrieves $PUB_{DH}$ and verifies that the nonce is incremented by 1. It generates a secret key b, and computes $B = g^b \bmod p$. Then it generates the secret key $K_S$ by $A^b \bmod p$. It encrypts the incremented nonce by $K_S$ and encrypts $B$ by $RND_S$. It forwards the packet to security management component. The security management component retrieves $B$ by decrypting with $RND_S$ and generates secret key $K_S$ by $B^a \bmod p$. It also retrieves the new incremented nonce by decrypting with $K_S$. It again increments the new nonce by 1 and encrypts with $K_S$. It then forwards the packet to the slave device. The slave device verifies that the nonce is again incremented by 1.

**Algorithm 4** Asymmetric key establishment

1: **procedure** ASYMKEYGEN

$SM:$ $\quad a = KeyGen(), A = g^a \bmod p$

$SM:$ $\quad P_{DH} = E(RND_S, (Pub_{DH}, (A), inc(NONCE_S)))$

$SM \rightarrow S:$ $\quad < P_{DH} >$

$S:$ $\quad < Pub_{DH}, (A), inc(NONCE_S) >= D(RND_S, P_{DH})$

$S:$ $\quad b = KeyGen(), B = g^b \bmod p$

$\quad\quad K_S = A^b \bmod p$

$S \rightarrow SM:$ $\quad < E(K_S, inc(NONCE_S)), E(RND_S, B) >$

$SM:$ $\quad B = D(RND_S, (E(RND_S, B)))$

$SM:$ $\quad K_S = B^a \bmod p$

$\quad\quad inc(NONCE_S) = D(K_S, (E(K_S, inc(NONCE_S))))$

$SM \rightarrow S:$ $\quad < E(K_S, inc(NONCE_S)) >$

$S:$ $\quad inc(NONCE_S) = D(K_S, (E(K_S, inc(NONCE_S))))$

2: **end procedure**

# 5. PROOF-OF-CONCEPT IMPLEMENTATION

We have implemented the deployment framework to verify the feasibility of our proposed scheme. The device deployment framework is implemented using four components, Employee

Management System (*EMS*), Commissioning Device (*HH*), Field Device (*FD*) and Security Manager (*SM*).

The overall packet transfer in this proof-of-concept implementation is presented in Figure 4 to make it easier for readers to visualize the framework implementation. The *EMS* component keeps the *APARAM* as secret. It encrypts the *APARAM* with the *EMS* public key and downloads it to the *ID card*. The *HH* component, takes this encrypted *APARAM* value once the employee verification is done and adds the configuration of the slave device. Then it encrypts the whole packet with the *EMS* public key. The *HH* also downloads the slave configuration file inside the device. The slave device takes the encrypted packet and adds a generated nonce and its identity. Then it encrypts the whole packet with the *EMS* public key and sends it to the next level of device. After receiving the packet, the master device adds its configuration data and encrypts the whole packet with the *EMS* public key. It also signs its identity and forwards the packet to the next forwarding device or the *EMS*. Using the private key of the *EMS*, the *EMS* can retrieve the forwarding device details, joining device details, the configuration of joining device and the *APARAM*.

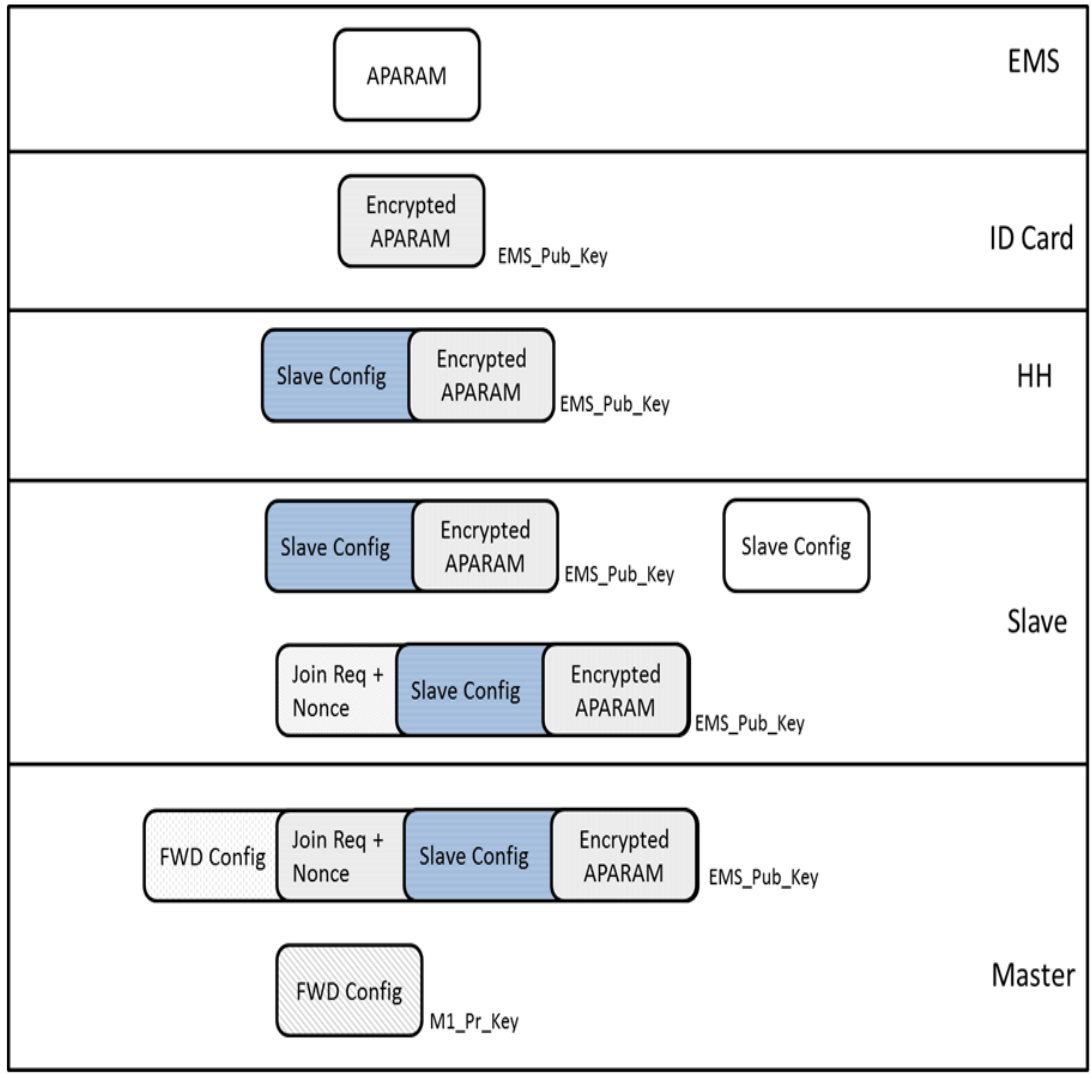

Figure 4.  Data Flow in Proof-of-concept Implementation

This implementation shows that the proposed framework is simple to implement. In our framework, we can use standardized encryption functions such as AES, 3DES or cipher block chaining libraries for encryption, decryption and signature verification. Therefore, this framework utilizes available standard security libraries for implementing those algorithms and this accelerates the implementation phase.

# 6. ASSESSMENT AND DISCUSSION OF THE DEVICE DEPLOYMENT FRAMEWORK FOR INDUSTRIAL PLANTS

In this paper, we have proposed a framework for efficient, user friendly device deployment reusing the concept of initial trust establishment. Our aim is to ensure that the entities in the industrial plants are communicating through a secure channel. In this section we will discuss whether this framework fulfils the objectives as mentioned earlier along with comparisons between different industry standard practices. We will also analyse the protocol using the AVISPA tool [35].

## 6.1. Framework performance comparison

As mentioned in Section 2, there is no automated workflow of initial credential distribution for industrial devices to the best of our knowledge. Hence, we will focus on the performance improvement through the proposed deployment framework compared to the industry current practices.

**Overview of different initial key distribution workflows in industrial plants:**

The initial key distribution in industrial plant is broadly categorized in seven categories [3]. These are:

(a) Master Device provides unique Symmetric Key for every device, (b) Master Device provides same Symmetric Key for all devices, (c) Master Device provides Public/Private key pair for Slave Device, (d) Device Manufacturer provides unique Symmetric Key for every device, (e) Device Manufacturer provides same Symmetric Key for all devices, (f) Device manufacturer provides Public/Private key pair, (g) Slave device provides Public/Private key pair. We summarize workflows for initial key distribution in Table 3.

We also define two broad categories of channels for key distributions. The first one is the *Trusted Channel* which is the medium where communicating parties are authenticated, though transmitted messages can be public. The second type of channel is the *Secured Channel* which is the medium where no one can listen to the exchanged messages except communicating parties.

Table 3.  Overview of initial key distribution workflow in industrial plants

| Approaches | Type of Channel | Property |
|---|---|---|
| Approach 1: Master device provides unique Symmetric Key for every device | Out-of-band | Secure channel |
| Approach 2: Master device provides same Symmetric Key for all devices | Out-of-band | Secure channel |
| Approach 3: Master device provides Public/Private key pair for Slave Device | Out-of-band | Secure channel for private key and Trusted channel for public key |

| Approach 4: Device Manufacturer provides unique Symmetric Key for every device | Out-of-band | Secure channel |
|---|---|---|
| Approach 5: Device Manufacturer provides same Symmetric Key for all device | Out-of-band | Secure channel |
| Approach 6: Device manufacturer Provides Public/Private key pair | Out-of-band | Trusted channel for public key |
| Approach 7: Slave device provides Public/Private key pair | Out-of-band | Trusted channel for public key |

**A comparison of different initial key distribution workflows for industrial plants:**

In all these seven approaches we mentioned, the public/private key or symmetric key is required to be installed in the device using an out-of-band mechanism. This requires a trusted, or trusted and secured channel. Table 4 presents a high level comparison between proposed method and other approaches for the following objectives.
Framework Objectives:

- Objective 1: Initial secret key never leaves the node
- Objective 2: System resilience
- Objective 3: Accountability for device configuration
- Objective 3: Ease of configuration
- Objective 5: Time to configure
- Objective 6: Ease of system deployment Security Objectives:
- Objective 7: Confidentiality
- Objective 8: Integrity
- Objective 9: Device Authentication

Table 4.  A comparison of workflows for initial credential distributions in industrial devices.

| Approach | Framework Objectives | | | | | | Security Objectives | | |
|---|---|---|---|---|---|---|---|---|---|
| | Obj 1 | Obj 2 | Obj 3 | Obj 4 | Obj 5 | Obj 6 | Obj 7 | Obj 8 | Obj 9 |
| Master Device provides Unique Symmetric Key for every device | No | High | Low | Medium | Low | Medium | Yes | Yes | Yes |
| Master Device provides same Symmetric Key for all device | No | Low | Low | Low | Medium | Medium | Yes | Yes | Yes |
| Master Device provides Public/Private key pair for Slave Device | No | High | Low | Medium | Low | Medium | Yes | Yes | Yes |
| Device Manufacturer Provides unique Symmetric Key for | No | High | Low | High | Medium | Low | Yes | Yes | Yes |
| every device | | | | | | | | | |
| Device Manufacturer provides same Symmetric Key for all device | No | Low | Low | High | Medium | Low | Yes | Yes | Yes |
| Device manufacturer provides Public/Private key pair | Yes | High | Low | High | High | Low | Yes | Yes | Yes |
| Slave device provides Public/Private key pair | Yes | High | Low | Low | High | Low | Yes | Yes | Yes |
| Initial Trust Establishment Framework (Proposed Idea) | Yes | High | High | Low | High | Low | Yes | Yes | Yes |

As we know that, during symmetric key distribution, there is a need for a trusted and secured channel where no one can listen when the initial key is going to be distributed. In an industrial plant, there might be many employees who will be handling the commissioning of devices. Therefore, when the device is configured for key management, then the secured and trusted channel is also being handled by different employees in the plant. If the secret key is required to be entered during commissioning, the key will be known to the employee who is configuring the device. For example, when an employee is commissioning/configuring 100 devices, there is a need to access 100 different secret symmetric keys for 100 devices. This affects the *initial secret key never leaves the node* property. Entering manually a symmetric key, which might be a 16 digit number, is an error prone and tedious job for the commissioning engineer. This reduces the *ease of configuration* of the system. In addition to it, the secret key is also getting revealed while entering the key during configuration. If a key is leaked in the network, it is difficult to find who has initiated the problem, as individual *accountability* is not tied with device configuration. To improve the *ease of configuration*, there is a probability of using the same initial bootstrapping key for all the devices in the network. However, this reduces the *resilience* of the system. If the same key is used to bootstrap all the devices in the network, then compromise of a single device will have high impact on the whole system. Therefore, approach 1 and 2 reveals the key but approach 1 has low *ease of configuration* and high *resilience*, whereas approach 2 has medium *ease of configuration* but low *resilience*. The *time to configure* property is also medium for both the approaches as it needs to configure security parameters during maintenance or replacement of devices. The problem of individual *accountability* of employees is also not solved, as we will not be able to identify who has commissioned the device. There is also a medium effort to set up a central security management component like Master Device which handles the security of large number of devices in industrial plants, which affects the *system deployment* property.

When public key cryptography is used, a trusted channel is created to transmit the public key. However, when the private/public key pair is generated from a central security server inside the plant, there is also a requirement of a secured channel to transfer the private key inside the device. Creating a secure channel to transmit the private key has similar usability issues similar to symmetric key distribution. Therefore, in approach 3, private key leaves the environment through an out-of-band channel and has low *ease of configuration*. The *time to configure* property is medium as it needs to configure security parameters during maintenance or replacement of device. The problem of individual *accountability* of employees is also not solved, as we will not be able to identify who has commissioned the device. There is also a medium effort to set up a central security management component like Master Device which handles the security of large numbers of devices in industrial plants, which affects the *system deployment* property.

On the other hand in approach 4 and 5, if the vendor puts the secret key in the device during manufacturing, the same key has to be transferred to the industrial plant through a secured channel. This requires that employee will know the secret key for commissioning the system. The *resilience* property will be affected the same way as in approach 1 and 2, if the same key is used for all devices. This approach improves the *ease of configuration* to an extent as the device is not required to be configured with a symmetric key during commissioning. However, this increases the *time to configure* property as during maintenance or replacement of a device, the device manufacturer is required to be contacted for acquiring new key pair for devices. There is also a high effort to set up the trusted and secured channel between manufacturer and industrial plant, which affects the *system deployment* property.

When public key cryptography is used and the manufacturers are responsible for generating public/private key pairs in approach 6, we can remove the secret handling by employees. This will not reveal the secret key and the public key mechanism will improve the *resilience*. However, the *time to configure* property will be high and there will be a high effort in *system deployment* to set up the trusted and secured channel between manufacturer and industrial plant.
In approach 7, the slave device itself is capable of generating a public/private key pair. This improves most of the properties but *accountability* of the commissioning engineer is not tied with this approach. Therefore, if the device is misbehaving then it is difficult to know who has configured the device and whether the configuration issues have created the problem. It also assumes that the slave device is computationally efficient to generate public/private key pairs by themselves.

In our proposed framework, the employee needs to swipe the *ID card* to the handheld terminal and provide authenticity. The employee is not required to enter any specific secret key for the device, instead the *ID card* is used in the same way it is used to access factory entry. The devices can present this trust information to receive the keys from the security management component. Therefore, in this framework, the initial security parameters do not get revealed to the employee who is configuring, instead the encrypted parameters are used to verify the authenticity of the device and the engineer. Once the device is authenticated by the employee management system, the configuration data and the related information is transferred to the security management component. Then the security management component becomes responsible for the key management of the whole network.

In our proposed framework, the key which is distributed based on the device capability of supporting encryption, is limited to only two communicating parties. Therefore, if the attacker can retrieve the key for a particular slave, it cannot compromise the entire system and communication. If the key of the master device is compromised, then the slave devices which are under the cluster of that particular master device will be compromised. However, it cannot compromise the other master devices in the network. When asymmetric key cryptography is used, compromise of one particular device cannot compromise the entire system.

Devices which are involved in data communication are commissioned by commissioning or maintenance personnel. The employees are the authorized persons to handle a device, therefore when the commissioning person places the device in the network; the trust parameter of the employee which is stored in the *ID card* is transferred to the device. When the device presents the configuration credentials, it also presents the encrypted employee trust. Commissioning engineers have sufficient experience to demonstrate that they know the safety regulations and machine directives to formally ″sign-off″ a commissioned plant. Hence, a commissioning engineer is trusted for the operational safety of a plant. Verifying that employee trust, the device can be authenticated that it is commissioned by an authorized person. This trust of the employee is transferred to the device only when the authorized commissioning or maintenance personnel swipes the employee identification card in card reader. Therefore, any other device which is not commissioned by authorized persons in the plant can easily be detected in this framework as only the employee can store this encrypted trust. In the future if the device has the capability to read the identity card, then the trust can be transferred directly without the need of an additional commissioning device.

In our framework we provide a mechanism which integrates the employee management system with the security management component for devices. The employee management system deals with the management of the employees who handle the device in a plant or organization. The security management component deals with the security of the devices in the plant. To configure a device, our framework requires that the employee swipes the *ID card* in an ID card reader, like a handheld commissioning device, and enters the configuration data. This procedure does not take extra time compared to the commissioning time without any security mechanism. This provides a user friendly procedure for the employees without accessing the secure data storage or manually entering the security related parameters.

This framework partially satisfies physical security where once an attack is detected, it can be tracked who has configured the device. In earlier approaches, there was no individual accountability. However, our proposal is highly dependent on an Employee Management System. This might affect the *ease of system deployment* as our method assumes that inside the plant there is a first level of access control and this component is used to securely store the employee access data. This is an additional requirement on current employee management systems. However, this will be a one-time activity and in most industrial plants, there exist a system for employee management.

## 6.2. Formal verification and validation of framework using AVISPA

In this section the results of formal verification of our proposed framework is presented to verify the correctness of the protocols. AVISPA (Automated Validation of Internet Security Protocols and Applications) [35, 36] is used for the analysis of large-scale Internet security-sensitive protocols and applications. To specify the security protocol and their properties, the HLSPL (High Level Protocols Specification Language) language is used. Protocols to be studied by the AVISPA tool should be specified in HLPSL and written in a file with the extension hlpsl. The HLPSL specification is translated into the Intermediate Format (IF) using a hlpsl2if translator. IF is a lower-level language than HLPSL and is read directly by the back-ends of the AVISPA Tool. The AVISPA Tool comprises four back-ends; OFMC (On the Fly Model Checker), CL-AtSe (Constraint Logic based Attack Searcher), SATMC (SAT based Model Checker), TA4SP (Tree Automata based on Protocol Analyzer). These back-ends are used to identify flaws in protocols. SPAN [37, 38] is a security protocol animator for AVISPA which is designed to help protocol developers in writing HLPSL specifications. A HLPSL specification is composed of three parts, namely a list of definitions of roles, a list of declarations of goals, and the read call of the main role.

Roles are used as independent processes and they have a name, receive information by parameters and contain local declarations. To formally verify the protocols used in our framework, we have used basic roles similar to our implemented version, Employee Management System (*EMS*), Handheld Device (*HH*) and Security Manager (*SM*). We also modelled *ID card* also as a role. For the sake of completion we have separated the Field Device component into Master Device (*M*) and Slave Device (*S*). Each basic role is independent from the others and has initial information. In our implementation each role contains local declarations, initialization and transitions. The transitions in a role are spontaneous actions enabled when the state predicates on the left-hand side are true. In our implementation, the sessions of the protocol is described as the composed role. In composed roles, the roles can execute parallel or sequentially.

In this paper, we have used both the OFMC and AtSe back-ends using SPAN to verify our protocol. In SPAN, CAS+ is used as a language. In CAS+, we declare the identifiers of the protocol from certain types, namely user (principal name), public key, symmetric key, function, number. The Table 5, summarizes the identifiers used to verify our proposed protocol.

Table 5. Identifier declaration.

| Type | Identifiers |
|------|-------------|
| User | *EMS, ID, HH, S, $M_1, M_2$, SM* |
| Number | *APARAM, CD, $NONCE_S$, $RND_s$, $RND_{SM}$, $RND_{M1}$, $RND_{M2}$* |
| Public Key | *$K_{pub}(EMS)$, $K_{pub}(SM)$, $K_{pub}(M_1)$, $K_{pub}(M_2)$* |
| Function | Increment |

When a protocol execution initiates, each principal needs initial knowledge to compose its messages. The identifiers in user category need to have the knowledge of data it uses for its protocol execution. The Table 6, captures the knowledge of each user in our implementation.

Table 6. Knowledge of User.

| User | Knowledge |
|------|-----------|
| *EMS* | *EMS, ID, HH, $M_1$, $M_2$, SM*, Increment, $K_{pub}(EMS)$, $K_{pub}(SM)$, $K_{pub}(M_1)$, $K_{pub}(M_2)$ |
| *ID* | *EMS, ID, $K_{pub}(EMS)$* |
| *HH* | *EMS, ID, HH, S*, Increment, $K_{pub}(EMS)$ |
| *S* | *EMS, ID, HH, S, SM*, Increment, $K_{pub}(EMS)$ |
| *$M_1$* | *EMS, M1, M2, SM*, Increment, $K_{pub}(EMS)$, $K_{pub}(SM)$, $K_{pub}(M_1)$, $K_{pub}(M_2)$ |
| *$M_2$* | *EMS, M1, M2, SM*, Increment, $K_{pub}(EMS)$, $K_{pub}(SM)$, $K_{pub}(M_1)$, $K_{pub}(M_2)$ |
| *SM* | *EMS, M1, M2, SM*, Increment, $K_{pub}(EMS)$, $K_{pub}(SM)$, $K_{pub}(M_1)$, $K_{pub}(M_2)$ |

The message section contains the core of the protocol specification. We use the message exchange algorithms as discussed in Section 4. We declare the goal of verification as secrecy of *APARAM*, whether *SM* and *S* can authenticate each other by $RND_S$ and $RND_{SM}$ respectively. Each role communicates with other roles through Dolev-Yao channels. In Dolev-Yao model, the adversary can overhear, intercept, and synthesize any messages. We have analysed our protocol with OFMC and ATSC.

The On-the-Fly Model-Checker OFMC builds the infinite tree of the problem in a demand-driven way. The state space is represented by different symbolic techniques. By using this, OFMC can detect attacks fast and prove the protocol is correct. The CL-based Model-Checker (CL-AtSe) is used to translate any protocol specification into a set of constraints. This is useful to find attacks on protocols [35]. The analysis with both the OFMC and ATSE shows that our proposed protocol has no security flaw that can be detected by AVISPA.

# 7. CONCLUSIONS AND FUTURE WORK

In this paper, we have presented a framework for industrial device deployment. We started by introducing the objectives of the device deployment framework. Then we presented our framework in detail and assessed the objectives of the device deployment framework. It is found that the device can be verified by the security management component once the commissioning engineer or maintenance engineer has established the initial trust by transferring the employee parameters to the device. The configuration parameters can also be downloaded during the initial trust establishment. Therefore, based on the cryptographic computational capability of the device, our proposed framework can support both symmetric and asymmetric key distribution. By reusing the initial trust establishment workflow, this framework simplifies the key distribution mechanisms and eliminates the need of prior sharing of secret parameters. The initial trust establishment phase does not require any unique secret for the device which is difficult to manage, rather the key distribution occurs from a central management component once the device can show that it has been commissioned by an authorized person. The authentication verification phase also provides a mechanism for the device to verify whether it is joining the intended network. We also logically segregate the security management for devices from the role of the commissioning engineer. Therefore, this framework provides a solution for the dynamic environment of employee roles in industrial plants. This framework is also adaptive where the devices do not have direct connectivity with the central security management or employee management system. Through the proposed authentication in direct and hierarchical topology, any device can be verified once initial trust has been established by the commissioning engineer.
As future work, we are planning to demonstrate the practicability of this framework with working devices in a plant.

## ACKNOWLEDGEMENTS

This work has been supported by the Swedish Knowledge Foundation (KKS) through ITS-EASY, Embedded Software and Systems Industrial Research School, affiliated with the School of Innovation, Design and Engineering (IDT) at Malardalen University (MDH, Vasteras, Sweden) as well as by the ABB Industrial Communication and Electronics Program
.